\documentstyle[preprint,prl,aps]{revtex}
\newcommand{\be}{\begin{equation}}
\newcommand{\ee}{\end{equation}}
\newcommand{\ba}{\begin{eqnarray}}
\newcommand{\ea}{\end{eqnarray}}
\newcommand{\ep}{\epsilon_{\mu \nu \alpha}}
\newcommand{\oh}{\displaystyle{\frac{1}{2}}}
\begin{document}
\draft
\title{Loop Corrections and Bosonization Formulae.}
\author{C.D.~Fosco$^a$\thanks{CONICET}
\\
C.~N\'u\~nez$^{b}$\thanks{CONICET} and F. A. Schaposnik$^b$\thanks{CICBA}
\\
{\normalsize\it
$^a$Centro At\'omico Bariloche,
8400 Bariloche, Argentina}\\
{\normalsize\it
$^b$Departamento de F\'\i sica, Universidad Nacional de La Plata}\\
{\normalsize\it
C.C. 67, 1900 La Plata, Argentina}}
\date{\today}
\maketitle
\begin{abstract}
We study the functional integrals that appear in a path integral 
bosonization procedure for more than two spacetime dimensions. Since
they are not in general exactly solvable, their evaluation by 
a suitable loop expansion would be a natural procedure, even if the
exact fermionic determinant were known. The outcome of our study 
is that we can consistently ignore loop corrections in the functional 
integral defining the bosonized action, if the same is done for the 
functional integral corresponding to the bosonic representation 
of the generating functional. If contributions up to some order $l$ in 
the number of loops are included in both integrals, all but the
lowest terms cancel out in the final result for the generating
functional.
\end{abstract}
\bigskip
\section{Introduction.}
The very useful property that the configurations of a given physical 
system can be equivalently described by different sets of variables has 
one of its more extreme manifestations in the bosonization procedure.
This was originally created to deal with some two-dimensional models, 
which could be described in terms of either fermionic or bosonic 
variables. Bosonization turned out to be not just a curiosity but
also a very useful tool indeed to understand and in some cases 
even to solve non-trivial interacting Quantum Field Theory models.
 
It is interesting to remark that there is no theoretical obstacle to 
the extension of this procedure to higher dimensions. Indeed, there has 
recently been some progress in the application, although in an 
approximated form, of the bosonization procedure to theories in more 
than two dimensions \cite{BQ}-\cite{LNS}, dealing with both the
Abelian and the non-Abelian cases. 

The essential difficulty which makes this extended bosonization procedure
non-exact is our inability to compute exactly a fermionic determinant 
in more than two spacetime dimensions. There is however another problem
which seems to call for additional approximations, even if the fermionic
determinant were exactly known. This is the fact that in order to
obtain the bosonized action one has to calculate a functional Fourier
transformation of the fermionic determinant. As the latter is in
general a non-local and/or non-polynomial function, it is in general
impossible to calculate that functional integral exactly, except for
simple situations (like the quadratic approximation for the fermionic
effective action). It is also possible to find the bosonized action  
for some non-trivial situations, like the non-Abelian case in three 
dimensions in a derivative expansion, by taking advantage of an 
underlying BRST symmetry. This symmetry however, is not powerful
enough in the Abelian case as to allow us to obtain the bosonized
action \cite{LGMNS}. 

One would expect on intuitive grounds that, once the fermionic
determinant is known, there should not be any physically relevant 
loop correction to perform. This is indeed what we will demonstrate
below, namely, no loop corrections are necessary beyond the (one-loop)
calculation involved in the fermionic determinant. This does not
mean that one {\em must not\/} calculate them, but rather that one
can consistently ignore them in the integrals over the bosonic fields
without affecting the exactness of the final result for the current
correlation functions in the bosonic approach. Moreover, we shall show
that if loops corrections are included, they do cancel in the final
result. 

The structure of this paper is as follows: In section 2 we explain
the mechanism to evaluate the integrals in a `minimal' way,
understanding by that that the minimum number (i.e., zero) of loops
has to be included in the path integrals over bosonic fields. 
We introduce the factors of $\hbar$ in order
to combine what is of the same order, and to separate what is
irrelevant to the physically meaningful results. Then  we explain 
how do the loop corrections cancel (in non-minimal approaches) and extend
our results to higher dimensional spaces. 
Section 3 presents the application of the previous results to a 
particular example, which consists of 
the Abelian case with the (log of the) fermionic determinant 
evaluated up to quartic order in the external field. We show 
explicitly the cancellation of one-loop diagrams.
Section 4 contains an independent justification of the `minimal'
or `classical' approach, and its application to the Abelian
and non-Abelian cases.

\section{Cancellation of Loop Corrections.}
For the sake of simplicity, we shall be first concerned with the 
Abelian case in $2+1$ dimensions. Different cases will be 
considered afterwards. 
Our starting point is the bosonic form for ${\cal Z}(s)$, the generating
functional of connected current correlation functions
\be
{\cal Z}(s) \;=\; \int {\cal D}A_\mu \, \exp \left[ - S_{bos}(A) -i \int d^3 x
\epsilon_{\mu \nu \lambda} s_\mu \partial_\nu A_\lambda \right]
\label{zbos}
\ee
where $s_\mu$ is an external source, $A_\mu$ is a bosonic gauge field, 
and $S_{bos}(A)$ is the `bosonic action', a functional of $A_\mu$ that
encodes all the fermionic current correlation functions in the bosonic
description. It is defined by a sort of functional Fourier transformation
of $Z(b)= \exp [-W(b)]$:
\be
\exp[-S_{bos}(A)]\;=\; \int {\cal D} b_\mu \, \exp \left[ - W(b) + i
\int d^3 x \epsilon_{\mu\nu\lambda} b_\mu \partial_\nu A_\lambda
\right]\;.
\label{sbos}
\ee

$W(b)$ is the generating functional of {\em connected\/} current
correlation functions. On the other hand, in the fermionic representation
${\cal Z}(s)$ is expressed by the functional integral
\be
{\cal Z} (s) \;=\; \int {\cal D} \psi {\cal D} {\bar \psi} \, 
\exp \left[- \int d^3 x {\bar \psi} (\not \! \partial + M + i \not \! s)
\psi \right]  \;.
\label{zfer}
\ee 

The connected correlation functions of the current $j_\mu (x) = 
{\bar \psi}(x)\gamma_\mu \psi (x)$ in the presence of the external 
source $s_\mu$ are
\be
\langle {j_\mu}_1 (x_1) \cdots {j_\mu}_n (x_n) \rangle_{conn} \;=\;
\frac{\delta}{i \delta {s_\mu}_1 (x_1)} \cdots 
\frac{\delta}{i \delta {s_\mu}_n (x_n)} W(s) \;.
\ee

In more than two spacetime dimensions, we do not know the exact
expression for ${\cal Z}(b)$ (and hence for $W(b)$). Even in
the hypothetical case of knowing the exact form of the fermionic
determinant, we should then have to confront the (perhaps more cumbersome)
task of calculating the functional integral over the auxiliary field
$b_\mu$ in (\ref{sbos}). And having thus obtained (either exactly,
or in some approximation) that functional integral, one should then
use the resulting bosonic action $S_{bos}$ in (\ref{zbos}) in order
to calculate correlation functions in the bosonic version of the
theory. This is again in general a non-trivial functional integral,
where there seems to be no hope for an exact evaluation.

It is the purpose of this section to clarify some issues related to the 
evaluation of the functional integrals over the fields $b_\mu$ and
$A_\mu$, appearing in equations (\ref{sbos}) and (\ref{zbos}), 
respectively. In particular, by keeping track of the dependence
on ${\hbar}$, we show that there is a consistency requisite for the
approximations done in the evaluation of those integrals, namely,
that both should be evaluated up to the same order in the number of
loops. This admits the `minimal' solution of using for both integrals  
the `tree' approximation. 
 
If we reintroduce the dependence on $\hbar$ in the fermionic
version (\ref{zfer}) of the generating functional of the full current
correlation functions, we see that it should be rewritten as
\be
{\cal Z} (s) \;=\; \int {\cal D}\psi {\cal D}{\bar \psi} \, 
\exp \left[- \frac{1}{\hbar} \int d^3 x {\bar \psi} (\not \! \partial 
+ M + i \not \! s) \psi \right]  \;.
\label{zferh}
\ee
This means that, for a given diagram, each fermionic line will have
a factor $\hbar$ attached, while each vertex will introduce a 
factor of $\hbar^{-1}$. 
As the functional integral (\ref{zferh}) only contains one-loop diagrams, 
with an equal number of fermion lines and vertices, all the factors
of $\hbar$ cancel out. Thus ${\cal Z}$ is independent of $\hbar$. 
When $\hbar$ is reintroduced into the game, one also modifies the
relation between $W$ and ${\cal Z}$
\be
\frac{1}{\hbar} W(b) \;=\; - \log Z(b) \;,
\label{wbh}
\ee
so that $W$ has the dimensions of an action. As, by counting
powers of $\hbar$, we have seen that ${\cal Z}$ is independent
of $\hbar$, we can use (\ref{wbh}) to make the statement that
$W(b)$ is of order $\hbar$. Namely,
\be
W(b) \;=\; \hbar \, {\cal W} (b)
\ee
where ${\cal W}$ is independent of $\hbar$. Of course we are
not saying more than the well-known fact that a 1PI diagram
with $L$ loops carries a factor of $\hbar^{L-1}$ (and $L=1$
in our case).

Now we deal with the functional integral over $b_\mu$ in 
(\ref{sbos}). From the previous review, we see that
after reintroducing $\hbar$, the proper expression for
that integral is
\be
\exp[- \frac{1}{\hbar}S_{bos}(A)]\;=\; \int {\cal D} b_\mu \, 
\exp \left[ - \frac{1}{\hbar} W(b) + i \int d^3 x 
\epsilon_{\mu\nu\lambda} b_\mu \partial_\nu A_\lambda \right] \;. 
\label{sbosh}
\ee

We now note that we have in (\ref{sbosh}) a functional integral
where $W(b)$ plays the role of an `action' for the field $b_\mu$.
By contrast with the usual situation, the `action' $W(b)$ is
of order $\hbar$. At this point, we should decide up to which
order in $\hbar$  we will work. We shall first develop what we
call the `minimal' solution, namely, the lowest approximation
which however yields an exact result for the correlation 
functions. Non minimal solutions will be discussed afterwards.

In this approximation, we want to know the result of the functional 
integral (\ref{sbosh}) up to order $\hbar$ (which is indeed the
order of $W$). As $W(b)$ is already of order $\hbar$, we only need to 
use the `classical' approximation, namely, 
\be
\exp[- \frac{1}{\hbar}S_{bos}(A)]\;\simeq\; 
\exp \left\{ - \frac{1}{\hbar} W[ {\hat b}(A)] 
+i \int d^3 x 
\epsilon_{\mu\nu\lambda} {\hat b}_\mu (A) \partial_\nu A_\lambda
\right\}
\;\equiv\;  
\exp[- \frac{1}{\hbar}S^{cl}_{bos}(A)]
\;,
\label{clas}
\ee
where ${\hat b}_\mu (A)$ is the solution to the equation
\be
\frac{\delta}{\delta b_\mu (x)} W({\hat b}) \,-\, i \epsilon_{\mu\nu\lambda}
\partial_\nu A_\lambda (x) \;=\; 0 \;. 
\label{sbcl}
\ee

It is clear that the `classical' bosonic action $S^{cl}_{bos}$ 
is of order $\hbar$. One should however not confuse our procedure
with a saddle point evaluation of the integral (which wouldn't be
in place here). It is more simple than that: In the functional integral
(\ref{sbosh}), $W$ plays the role of an `action' for the field $b_\mu$,
$i \epsilon_{\mu\nu\lambda}\partial_\nu A_\lambda$ is a `source' for
$b$, and $S_{bos}$ is the would be generating functional of connected
correlation functions of $b$ (except for a trivial operator acting
on each leg). We know on general grounds that  $S_{bos}$ is related
to the corresponding effective action by a simple Legendre
transformation. To order $\hbar$, this effective action is just the
`action' $W(b)$, and the Legendre transformation is accomplished by
equations (\ref{clas}) and (\ref{sbcl}).

Inserting (\ref{sbcl}) into (\ref{zbos}), 
\be
{\cal Z}(s) \simeq 
\int {\cal D}A_\mu \, \exp \left[ - \frac{1}{\hbar} S^{cl}_{bos}(A) -
i \int d^3 x \epsilon_{\mu \nu \lambda} s_\mu 
\partial_\nu A_\lambda \right]
\label{zbcl}
\ee
where it becomes evident that, as far as order $\hbar$ results are
concerned, we can once more calculate the integral in the tree
level approximation. This is tantamount to evaluating the exponent at its
extreme value
\be
{\cal Z}(s) \;\simeq\; 
\exp \left[ - \frac{1}{\hbar} S^{cl}_{bos}({\hat A}(s)) -
i  \int d^3 x \epsilon_{\mu \nu \lambda} s_\mu 
\partial_\nu {\hat A}_\lambda (s) \right]
\label{zcl}
\ee
where ${\hat A}(s)$ satisfies 
\be
\frac{\delta}{\delta A_\mu (x)} \, 
S^{cl}_{bos}({\hat A}(s)) +
i  \epsilon_{\mu \nu \lambda} 
\partial_\nu  s_{\lambda}  \;=\; 0 \;.
\label{sacl}
\ee 
On the other hand, we can now check whether the two-step approximation
we have done, consisting of deriving first a bosonized action valid to
order $\hbar$ and then using this action to calculate ${\cal Z}(s)$
to the same order makes sense. We recall that ${\cal Z}(s)$ in
(\ref{zcl}) can be written as ${\cal Z}(s) = \exp [-\frac{1}{\hbar} 
W(s) ]$, where $W$ is of order $\hbar$. But we have calculated the same
object to the same order. Thus (\ref{zcl}) must be exact.
Indeed, this is so by virtue of the simple fact that what we have done in
this two-step procedure is nothing more than an iterated Legendre 
transformation on $W$, which is exactly involutive, and thus comes back
to $W$ after performing it twice. The first Legendre transformation
starts from $W(b)$ and goes to $S^{cl}_{bos}(A)$ (where $A$ is simply
related to the derivative of $W$ with respect to $b$); and the second
one starts from $S^{cl}_{bos}(A)$ and goes to a function of $s$ 
(related to the derivative of $S$ with respect to $A$),
which because of the involutive property is
\be
S^{cl}_{bos}({\hat A}(s)) \,+\,
i \int d^3 x \epsilon_{\mu \nu \lambda} s_\mu 
\partial_\nu {\hat A}_\lambda (s) \;=\; W(s) \;.
\ee 

It is worth remarking that neither the functional integral over $b_\mu$, nor
the one over $A_\mu$ has been exactly calculated, but we can approximate
them in a synchronized way as to preserve the result for the fermionic
determinant. Of course one might try to evaluate both integrals
exactly, but in the end no improvement upon the previous results
would be obtained. Moreover, our `minimal' approach assures that the
only true `quantum' corrections come from the fermionic determinant,
and no extra loops have to be computed if that object has already been
calculated. Note also that we avoid in this way a potentially dangerous
situation: Assume that we have for the exact fermionic determinant
vertices which are non-renormalizable. If loops have to be calculated
one should face the problem of making sense of a non-renormalizable 
theory, whereas this problem does not arise in our approach. Of course
there are also practical advantages, since we just have to calculate
tree diagrams in order to obtain the bosonized action, and since
the latter has to be used in the tree approximation, {\em this is
in fact the effective (1PI) action \/}.

We shall now elaborate upon the problem of including loops and
how do they cancel when both integrals are calculated.  

It should come as no surprise that some cancellation between
the loops corresponding to the two integrals should occur,
if one realizes that the input of the bosonization procedure is $W(b)$, 
a one-loop object, and the outcome is a bosonic representation of
the same object: $W(s)$. 

It is straightforward to put that cancellation in a more
evident fashion. We start from the integral over $b_\mu$ of
equation (\ref{sbos}), where we perform a shift from
$b_\mu$ to $\beta_\mu$: 
\be
b_\mu \;=\; {\hat b}_\mu (A) \,+\, \beta_\mu
\ee
where ${\hat b}(A)$ depends on $A$ in a generally complicated way
since it is given by (\ref{sbcl}). Then we can write
\be
\exp [-S_{bos} (A)] \;=\; \exp [ -{\hat S}_{bos}(A)
- \sigma (A) ] 
\label{shf1}
\ee
where 
\be
\exp [ -\sigma (A) ] \;=\; \int {\cal D}\beta \, 
\exp \left\{ - [ W({\hat b}_A + \beta)\;-\; W({\hat b}_A) ] + 
i \int d^3x \beta_\mu 
\epsilon_{\mu\nu\lambda}\partial_\nu A_\lambda \right\}
\label{dsig}
\ee
where ${\hat S}(A) \equiv S_{bos}^{cl}(A)$, as given by (\ref{clas}).
We note that there is no linear term in $\beta$ if the exponent
on the rhs of (\ref{shf1}) is expanded in powers of $\beta$. 
Analogously, we shift now the integration variable $A_\mu$ in 
the integral yielding the generating functional in the bosonic
representation. Now the shift is defined by
\be
A \;=\; {\hat A}_s + \alpha
\ee
where ${\hat A}_s$ depends on the source $s_\mu$, and is determined
by (\ref{sacl}). This leads to the exact relation
$$
e^{-W(s)} \;=\; exp \left[ - {\hat S}_{bos}({\hat A}_s) -
i \int d^3x s_\mu \epsilon_{\mu\nu\lambda}\partial_\nu 
{\hat A}_\lambda \right]
$$
\be
\int {\cal D}\alpha \exp \left\{ - [ {\hat S}_{bos}({\hat A}_s +
\alpha ) - {\hat S}_{bos}({\hat A}_s) ] - \sigma ({\hat A}_s
+\alpha ) - i \int d^3x \alpha_\mu \epsilon_{\mu\nu\lambda}
\partial_\nu s_\lambda \right\} \;. 
\ee
On the other hand, we know by the involution of the Legendre 
transformation that
\be
exp \left[ - {\hat S}_{bos}({\hat A}_s) -
i \int d^3x s_\mu \epsilon_{\mu\nu\lambda}\partial_\nu 
{\hat A}_\lambda \right]\;=\; \exp [-W(s)] \;,
\label{char}
\ee
thus we conclude that the integrals over the fluctuations
$\alpha$ and $\beta$ satisfy~\footnote{Note that $\sigma$ is
defined through an integration over $\beta$ in (\ref{dsig}).} 
the relation
\be
\int {\cal D}\alpha \exp \left\{ - [ {\hat S}_{bos}({\hat A}_s +
\alpha ) - {\hat S}_{bos}({\hat A}_s) ] - \sigma ({\hat A}_s
+\alpha ) - i \int d^3x \alpha_\mu \epsilon_{\mu\nu\lambda}
\partial_\nu s_\lambda \right\} \;=\;1 \;,
\label{can}
\ee
and this is the identity which, if expanded in loops, shows the
order by order cancellation.

Let us consider an extension of this three dimensional Abelian case to
higher dimensional spaces. In the $d$-dimensional case, 
the Lagrange multiplier
is a Kalb-Ramond field with $d-2$ indices, the partition functional reads,
\be
Z[s]= \int DA_{\mu_1.....\mu_{d-2}~} Db_{\mu} e^{(- W[b] +\frac{i}{2}\int
d^3 x A(f[b] -f[s]))}
\label{f}
\ee
where we have used the notation 
\be
A(f[b]-f[s]) = \epsilon_{\mu_1 ..... \mu_d}
(f_{\mu_1 \mu_2}[b] -f_{\mu_1 \mu_2}[s])A_{\mu_3 ... \mu_d}
\ee

As we have done before, we shift
\be
b_{\mu}= {\hat b}_{\mu}[A]  + \beta_{\mu}
\ee
with ${\hat b}_\mu [A]$ determined from the analogous of equation 
(\ref{sbcl}).
\be
\frac{\delta}{\delta b_{\mu_1} (x)} W({\hat b}) \,-\, i 
\epsilon_{\mu_1 ... \mu_d}
\partial_{\mu_2} A_{\mu_3 ... \mu_d} (x) \;=\; 0 \;. 
\label{sbcl2}
\ee
and then change the variables in the Kalb-Ramond field,
\be
A_{\mu_3 ... \mu_d}={\hat A}_{\mu_3 ... \mu_d}[s] + \alpha_{\mu_3 ...
\mu_d}
\ee
with ${\hat A}[s]$ obtained from
\be
\frac{\delta}{\delta A_{\mu_3 ... \mu_d} (x)} S_{cl}^{bos}({\hat A}) 
\,-\, i 
\epsilon_{\mu_1 ... \mu_d}
\partial_{\mu_1} s_{\mu_2} (x) \;=\; 0 \;. 
\label{sbcl3}
\ee
It is straightforward to obtain analogous 
expressions to eqs.(\ref{char})-(\ref{can}), which 
shows the order by order cancellation.
We show explicitly in the example of 
the next section how does this cancellation (in $d=3$) works at the
one-loop order.

\section{Application to the Abelian case in the quartic approximation.}
We shall apply here the minimal approach to the construction of the
bosonized action $S_{bos}(A)$ for the case of a massive fermionic field
in $3$ Euclidean dimensions, with the assumption that $W(b)$ 
($=-\log \det(\not \! \partial + i \not \! b + M )$) has been
evaluated up to order $4$ in the external field $b_\mu$. This is
a non-trivial addition to the already studied case of the
quadratic~\cite{BFO} approximation, where the problem we are now dealing 
with was absent, since the integrals were Gaussian. Moreover, the
results we will obtain may be relevant not only to the bosonization
of the four-current correlation function, but also for the case
of the two-point function in the presence of an external source.
It also shows clearly the interplay between the approximation
of retaining terms with up to four $b's$ and the minimal approximation
for the integral over that field. 

To begin with, we note that the most general form (in coordinate
space) for the functional $W(b)$ in the case at hand is 
$$
W(b)\;=\;\oh \int d^3x_1 d^3x_2 \, W^{(2)}_{\mu_1\mu_2} (x_1,x_2) 
b_{\mu_1}(x_1) b_{\mu_2}(x_2) 
$$
\be
+\;\frac{1}{4!} \int d^3x_1  d^3x_2 d^3x_3 d^3x_4 \, 
W^{(4)}_{\mu_1\mu_2\mu_3\mu_4} (x_1,x_2,x_3,x_4) 
b_{\mu_1}(x_1) b_{\mu_2}(x_2) b_{\mu_3}(x_3) b_{\mu_4}(x_4) \;,
\label{wfour}
\ee
where both $W^{(2)}_{\mu_1\mu_2} (x_1,x_2)$ and 
$W^{(4)}_{\mu_1\mu_2\mu_3\mu_4} (x_1,x_2,x_3,x_4)$ are symmetrical 
functions under a simultaneous permutation of their space arguments 
and indices. Note that the term linear in $b$ vanishes, as usual, 
and the possible term with three $b$'s is absent because we are 
dealing with the Abelian case~\footnote{We should need an $W^{(3)}$ 
with the properties of being symmetric, transverse, and 
parity-violating. This cannot be built in three dimensions.}.
In order to find the bosonized action, we have first to solve
equation (\ref{sbcl}), which in terms of expansion (\ref{wfour})
may be written in the following form
$$
b_\mu (x)\;=\; i \int d^3y_1 G_{\mu\nu_1}(x,y_1) f_{\nu_1}(y_1)
$$ 
\be
-\,\frac{1}{3 !} \int d^3y_1 d^3y_2 d^3y_3 d^3y_4\,
G_{\mu\nu_1}(x,y_1) W^{(4)}_{\nu_1\nu_2\nu_3\nu_4}(y_1,y_2,y_3,y_4) 
b_{\nu_2}(y_2) b_{\nu_3}(y_3) b_{\nu_4}(y_4) 
\label{cleq}
\ee
where $f_\mu (x) = \epsilon_{\mu\nu\lambda}\partial_\nu A_\lambda$,
and $G_{\mu\nu}(x,y)= [W^{(2)}]^{-1}_{\mu\nu}(x,y)$.
We have on purpose stopped the expansion up to order three in the
field $A_\mu$. The reason is that when this expansion is inserted into
the expression for the bosonic action  (\ref{clas}), higher order
terms would give for the bosonic action terms with more than four
$A$'s, which would correspond to correlation functions of more than
four currents. On the other hand, we only know $W$ in the quartic
approximation, so the inclusion of those higher order terms would
give unreliable results. The bosonized action that follows from
expansion (\ref{cleq}) is
$$
S_{bos}^{cl} \;=\; \oh \int d^3x_1 d^3x_2 
[W^{(2)}]_{\mu_1\mu_2}^{-1} (x_1,x_2) 
f_{\mu_1}(x_1) f_{\mu_2}(x_2) 
$$
$$
+\,\int d^3x_1 d^3x_2 d^3x_3 d^3x_4 
d^3y_1 d^3y_2 d^3y_3 d^3y_4 
W^{(4)}_{\mu_1\mu_2\mu_3\mu_4}(x_1,x_2,x_3,x_4) 
$$
$$
[W^{(2)}]^{-1}_{\mu_1\nu_1} (x_1,y_1) 
[W^{(2)}]^{-1}_{\mu_2\nu_2} (x_2,y_2) 
[W^{(2)}]^{-1}_{\mu_3\nu_3} (x_3,y_3) 
[W^{(2)}]^{-1}_{\mu_4\nu_4} (x_4,y_4) 
$$
\be
f_{\nu_1}(y_1)f_{\nu_2}(y_2)f_{\nu_3}(y_3)f_{\nu_4}(y_4)
\label{rsbs}
\ee
which contains only terms up to order four in $A$, as promised.
It should become evident from the previous equation why the 
previous known results using the quadratic approximation for $W$ 
yielded the exact result for the two-current correlation
function: In the quadratic approximation no loops are possible,
and the result for the bosonized action is just the first term 
on the rhs of (\ref{rsbs}). Moreover, we can also affirm that
the result (\ref{rsbs}) yields the exact four-current
correlation function.

As a consistency check, we use the classical approximation to 
evaluate the functional integral over $A_\mu$. Solving
equation (\ref{sacl}) for $S_{bos}$ found in (\ref{rsbs})
to determine $A_\mu$ as a function of the source 
$s_\mu$ (up to order three in the source) yields
$$
f_\mu (x) \;=\; -i \int d^3y_1 W^{(2)}_{\mu\nu_1}(x,y_1)
s_{\nu_1}(y_1) 
$$ 
\be
-\; \frac{i}{3 !} \int d^3y_1 d^3y_2 d^3y_3 
W^{(4)}_{\mu{\nu_1}{\nu_2}{\nu_3}}(x,y_1,y_2,y_3)
s_{\nu_1}(y_1) s_{\nu_2}(y_2) s_{\nu_3}(y_3)\;.
\ee
And inserting this in the rhs of (\ref{zcl}) produces
the result
\be
{\cal Z}(s)\;=\; e^{- W(s) }
\ee
with
$$
W(s)\;=\;\oh \int d^3x_1 d^3x_2 \, W^{(2)}_{\mu_1\mu_2} (x_1,x_2) 
s_{\mu_1}(x_1) s_{\mu_2}(x_2) 
$$
\be
+\;\frac{1}{4!} \int d^3x_1  d^3x_2 d^3x_3 d^3x_4 \, 
W^{(4)}_{\mu_1\mu_2\mu_3\mu_4} (x_1,x_2,x_3,x_4) 
s_{\mu_1}(x_1) s_{\mu_2}(x_2) s_{\mu_3}(x_3) s_{\mu_4}(x_4) \;,
\ee
which is exactly equal (but now as a function of $s$ rather
than $b$) to the original assumption (\ref{wfour}) for the fermionic 
determinant in the presence of the external field. This confirms
our statement that the procedure we have followed introduces no
errors in the final answer beyond the ones involved in the
approximation of the determinant.

We now show the meaning of the cancellation of one-loop diagrams, 
as an illustration of the general result presented in the previous section.
We shall of course take into account that now we are dealing with
an approximation to the exact determinant because we have truncated the
expansion at the quartic term, so that the cancellation will show
up to this order.
It is a straightforward exercise to show that, if (\ref{can}) 
is expanded to one-loop order, we obtain the relation
\be
\det[ \frac{\delta^2{\hat S}({\hat A}_s)}{\delta A_\mu(x) 
\delta A_\nu(y)} ] \;\;
\det[ \frac{\delta^2{W}({\hat b}_{{\hat A}_s})}{\delta b_\mu(x) 
\delta b_\nu(y)} ]\;=\; 1 \;.
\ee
>From the previous example, we can of course extract the 
values of the two functional derivatives. It is 
simpler for the case of the derivatives with respect to
$A_\mu$ to consider the derivatives with respect to
$f_\mu = \epsilon_{\mu\nu\lambda}\partial_\nu A_\lambda$.
At the end, the determinants will only differ in the 
determinant of a field independent operator.
$$
\frac{\delta^2 {\hat S} ({\hat A}_s) }{\delta f_{\mu_1}(x_1) 
\delta f_{\mu_2}(x_2)} \;=\;  [W^{(2)}]^{-1}_{\mu_1\mu_2}
(x_1,x_2) \;-\; \oh  \int d^3z_1 d^3z_2 d^3y_1 d^3y_2 
[W^{(2)}]^{-1}_{\mu_1\rho_1}(x_1,z_1) 
[W^{(2)}]^{-1}_{\mu_2\rho_2}(x_2,z_2)
$$
$$
W^{(4)}_{\rho_1\rho_2\nu_1\nu_2}
(z_1,z_2,y_1,y_2) s_{\nu_1}(y_1) s_{\nu_2}(y_2)
$$
\be
\frac{\delta^2{W} ({\hat b}_{{\hat A}_s})}{\delta b_{\mu_1}(x_1) 
\delta b_{\mu_2}(x_2)} \;=\; W^{(2)}_{\mu\nu}(x_1,x_2)
\;+\; \oh \int d^3y_1 d^3y_2 W^{(4)}_{\mu_1\mu_2\nu_1\nu_2}
(x_1,x_2,y_1,y_2) s_{\nu_1}(y_1) s_{\nu_2}(y_2)\;.
\ee
The evaluation of these determinant up to the order we are
dealing with yields
$$
\log \det[ \frac{ \delta^2{\hat S} ( {\hat A}_s )  }{\delta A_{\mu_1}(x_1) 
\delta A_{\mu_2}(x_2) }] \;= 
$$
$$ -\oh {\rm Tr} 
\left\{\int d^3z_1 d^3y_1 d^3y_2 
[W^{(2)}]^{-1}_{\mu_1\rho_1}(x_1,z_1) 
W^{(4)}_{\rho_1\mu_2\nu_1\nu_2}
(z_1,x_2,y_1,y_2) s_{\nu_1}(y_1) s_{\nu_2}(y_2) \right\}
$$
$$
\log \det[\frac{ \delta^2 W ( {\hat b}_{{\hat A}_s} ) }{\delta b_{\mu_1}
(x_1) \delta b_{\mu_2}(x_2) } ] \;=\;
$$
\be
+\oh {\rm Tr} 
\left\{ \int d^3z_1 d^3y_1 d^3y_2 
[W^{(2)}]^{-1}_{\mu_1\rho_1}(x_1,z_1) 
W^{(4)}_{\rho_1\mu_2\nu_1\nu_2}
(z_1,x_2,y_1,y_2) s_{\nu_1}(y_1) s_{\nu_2}(y_2)
\right\}
\ee
where the cancellation becomes evident.

\section{A `classical' approach to Bosonization}

\subsection{Abelian Case}
In section II it was shown that the relation between the bosonic Action
for $A_{\mu}$, and the corresponding one for the $b_{\mu}$ field is
given by
$$
S_{bos}[A] =  W[b] - i\int d^{3}x \epsilon_{\mu\nu\alpha} b_{\mu}
\partial_{\nu} A_{\alpha}
$$
where the $b_{\mu}$ field has to be evaluated over the solution to the
equation
\be
\frac{\delta W[b]}{\delta b_{\mu}} - i
\epsilon_{\mu\nu\alpha}\partial_{\nu} A_{\alpha} \;=\;0 \;.
\ee

One possible way to mimic this construction would be to start from the
generating functional of connected current correlation functions, $W[s]$,
which is, of course, independent of  the `dynamical' field $b_{\mu}$.
We see that due to the gauge invariance of $W[s]$ under transformations
of $s$, we can write
\be
W[s_{\mu} + b_{\mu}]\vert_{F_{\mu\nu}[b]=0} = W[s_{\mu}]
\label{a}
\ee

We can represent the zero curvature condition using a Lagrange multiplier
field $A_{\mu}$ and rewrite eq.(\ref{a}),

\be
W[s]= {\cal W}[s, A, b]\vert _{A, b}
\label{33}
\ee
where we have defined,
\be
{\cal W}[s, A, b] = W[s + b] -
 i\int d^{3}x \ep A_{\mu}\partial_{\nu} b_{\alpha}.
\label{c}
\ee
Eq.(\ref{33}) means that $W={\cal W}$ when the fields $A, b$ 
are eliminated by using their equation of motion.
Indeed, if the equation for $A$ is used first,
\be
\frac{\delta W[s]}{\delta A_{\mu}} =0 \to \ep \partial_{\nu} b_{\alpha}=0
\label{d}                                                             
\ee
we see that $b$ can only be a pure gradient, gauge invariance of $W$
is used to prove (\ref{33}). If, in turn, the equation of motion of $b$
 were
used first, one would recover the minimal bosonized 
action of the previous section.

\subsection{Non-Abelian Case}
We will proceed along similar lines to those of the previous (Abelian) 
case. We start by writing the functional $W[s]$ as,
\be
W[s]=W[b] \vert_{F_{\mu\nu}[s] = F_{\mu\nu}[b]}
\label{1}
\ee
The condition $F_{\mu\nu}[s] = F_{\mu\nu}[b]$ implies, on a particular
section, $b_{\mu}=s_{\mu}$, as can be seen in 
\cite{LGMNS}
It will be useful to  rewrite eq (\ref{1}) defining 
as in the previous subsection,
$$
{\cal W}[s, A, b]\vert_{A,b}= W[s] 
$$
and
\be
{\cal W}[s, A, b]=W[b] +\frac{i}{2}
\int d^{3}x \ep ((A-b)_{\mu}F_{\nu\alpha}[b] +
(s-A)_{\mu}F_{\nu\alpha}[s] - 2\bar{c_{\mu}}D_{\nu}[b]c_{\alpha}) \;.
\label{2}
\ee
Here we have introduced the fields $A_{\mu}$ transforming as a vector
under gauge transformations, $\bar{c_{\mu}}$ and $c_{\mu}$ a pair of
anticommuting ghost, transforming covariantly under gauge transformations.
We have introduced these fields in a way reminiscent to that of ref. 
\cite{LGMNS}.
The purpose of writing eq.(\ref{2}) is to have an analogous identity as
the one derived in the Abelian case, in a way such that, after equations
of motion for the `dynamical' fields are used, eq.(\ref{2}) holds.

The ghost term in the previous equation may be thought of as a coming from
a (partial) gauge fixing for the topological gauge invariance 
$b_{\mu}\to b_{\mu} + \epsilon_{\mu}$ present in $W[s]$, since it
is in fact {\em independent} of $b_\mu$. It is partial because there
remains a non-Abelian gauge invariance (of the usual kind), 
since $b_\mu$ is only fixed up to gauge transformations. The
Fadeev-Popov like term is the one just needed for the measure of the 
functional integral to be well-defined, we add this term in the classical 
action (\ref{2}) in a way reminiscent to that in the quantum theory (see
ref. \cite{LGMNS}). It has to be included if the quantum theories
following from different gauge fixings are to be equivalent. 

Differentiation of both sides of (\ref{2}) gives the eqs. of motion for
the `dynamical' fields. Indeed, differentiation with respect to $b_{\mu}$
gives,
\be
\frac{\delta W[b]}{\delta b_{\mu}} -2i\ep (\frac{1}{2} F_{\nu\alpha}[b]
+D_{\nu}[b](A-b)_{\alpha} + [\bar{c_{\alpha}},c_{\nu}])=0,
\label{3} 
\ee
with respect to $A_{\mu}$
\be
\ep F_{\nu\alpha}[b] = \ep F_{\nu\alpha}[s]
\label{4}
\ee
and with respect to the ghost fields,
$$
\ep D_{\nu}[b]\bar{c_{\alpha}}=0
$$
\be
\ep D_{\nu}[b]c_{\alpha}=0
\label{44}
\ee

Inserting the solution of eqs. (\ref{4}-{44})
$b_{\mu}=s_{\mu}$, $\bar{c_{\mu}}=c_{\mu}=0$ in eq.(\ref{3}) 
we obtain,
\be
\frac{\delta W[s]}{\delta s_{\mu}} -i\ep(1/2 F_{\nu\alpha}[s]
+ D_{\nu}[s](A-s)_{\alpha} )=0,
\label{5}
\ee
This is the analogous of the equations previously obtained, to which it reduces
in the Abelian case.

The outcome of this discussion is that, by construction, one can
indeed start from the functional $W[b]$, calculated in some
approximation, and define {\em classically} a bosonized form.
The procedure is just to evaluate ${\cal W}$ on the equations of
motion for $b_\mu$, what allows one to get an expression
that depends on $A$, the ghosts, and the source. Taking derivatives 
with respect to the source the bosonization rules are derived.
As the whole procedure stems from an exact classical relation,
it is evident that no quantum corrections are required. However,
the derivation of the exact classical bosonized action shall
require classical perturbation theory, since the equations that
determine this functional are in general non-linear (except
when the quadratic approximation for $W[s]$ is used).

\underline{Acknowledgments}: 
C.N. would like to acknowledge the warm hospitality at Instituto Balseiro
where part of this work was one. 

\end{document}